\documentclass[11pt,a4 paper]{article}
\usepackage{amssymb}
\usepackage{amsmath}
\usepackage[dvips]{graphicx}
\begin{document}
\thispagestyle{empty}
\noindent {\textbf{\Large New agegraphic dark energy in Brans-Dicke theory with logarithmic form of scalar field}}

\vspace{1cm}

\noindent  \textbf{Pankaj Kumar\;$^{2}$, \;\textbf{C. P. Singh\;\footnote{Corresponding author}}}\\

\vspace{0.5cm}

\noindent{ $^{1,2}$Department of Applied Mathematics,\\
 Delhi Technological University \\
 Bawana Road, Delhi-110 042, India.}\\
\texttt{ $^1$cpsphd@rediffmail.com  \\
$^2$pankaj.11dtu@gmail.com}\\

\vspace{1.5cm}
\noindent {\textbf{Abstract}} Recently, the current authors (arXiv:gr-qc/1609.01477) have proposed and analyzed in detail the logarithmic form of Brans-Dicke scalar field $\phi$ as $\phi \propto ln(\alpha+\beta a)$, where $\alpha$ and $\beta$ are positive constants, to alleviate the problems of interacting holographic dark energy models in Brans-Dicke theory. In this paper, the cosmological evolution of a new agegraphic dark energy (NADE) model within the framework of Friedmann-Robertson-Walker Universe is analyzed with the same form of scalar field in Brans-Dicke theory. We derive the equation of state parameter $w_D$ and deceleration parameter $q$ of NADE model. It is observed that $w_D\rightarrow -1$ when  $a\rightarrow \infty$, i.e., the NADE mimics cosmological constant in the late time evolution. Indeed, due to the assumption of logarithmic form of Brans-Dicke scalar field the NADE in Brans-Dicke theory behaves like NADE in general relativity in the late time evolution. The NADE model shows a phase transition from matter dominated phase in early time to accelerated phase in late time. We further extend NADE model by including the interaction between dark matter and NADE. In this case, $w_D$ definitely crosses the phantom divide line ($w_D=-1$) in the late time evolution. The phase transition from matter dominated to NADE dominated phase may be achieved at early stage in interacting model. Further, we show that the interacting NADE model resolves the cosmic coincidence problem as the energy density ratio may evolve sufficiently slow at present.\\

\noindent Keywords- New agegraphic dark energy, Brans-Dicke theory, Coincidence problem.
\pagebreak

\pagestyle{myheadings}
\section{Introduction}
\indent The cosmological observations like type Ia supernovae \cite{riess,perl} have provided the convincing evidence that our Universe is dominated by two dark components containing dark matter (DM) and dark energy (DE). The nature of DM and DE is one of the most important issues today in modern cosmology. The dark matter plays a role in explaining galactic curves and large scale structure where as dark energy explains the present cosmic accelerated expansion of the Universe. In the literature, the accelerating cosmologies have been described by two ways : dark energy content and modified gravity. In DE models, the accelerated expansion has been explained by modifying the right-hand side of the Einstein equations, i.e., by modifying the energy-momentum tensor. The modified gravity theory corresponds to modifying the left-hand side of Einstein equations, i.e., by modifying the geometric part. We refer \cite{sami,bamba} for detail study of DE models and modified gravity. \\
\indent However, we are still far from a compelling theory which could explain not only the recent accelerated expansion but also the primordial cosmology, structure formation, dark matter problem etc. without facing any problem. Whatever the way we choose to explain these problems, the quantum effects should be incorporated to come up with an accurate theory of gravity. The gravity theory which incorporates the principles of quantum mechanics, known as quantum theory of gravity. Although, the quantum gravity is not a well established theory, there are some concepts which have been proposed based on the principles of quantum gravity. In this context, the holographic dark energy (HDE) and agegraphic dark energy (ADE), which contain some significant properties of the quantum gravity, have been proposed as a candidate of DE to explain the recent accelerated expansion of the Universe. These DE models are able to present a unified model of the Universe and explain some other problems of the modern cosmology like coincidence problem \cite{nojiri}.\\
\indent The HDE models are based on the holographic principle \cite{hooft} and have been studied extensively in the literature \cite{li,pav, set,ban,LXU,Shey,LI,ghaf} . On the other hand, the ADE models are based on the quantum fluctuations of the space-time. Using the concept of quantum fluctuations of space-time, K$\acute{a}$rolyh$\acute{a}$zy and his collaborators \cite{karol,Karol,KArol} have observed that the distance $t$ in Minkowaski space-time can not be known to a better accuracy than
\begin{equation}
\delta t=\gamma\; t_{p}^{2/3}\; t^{1/3},
\end{equation}
where $t_p$ denotes reduced Planck time and $\gamma$ is a dimensionless constant of order unity. We assume $\hbar=k_b=c=1$ throughout in this paper. Maziashvili \cite{maz,Maz} has argued that the quantum energy density of the metric fluctuations of Minkowaski space-time can be estimated using the relation (1) with the time-energy uncertainty relation. Maziashvili \cite{maz,Maz} and Sasakura \cite{sasa} have independently obtained the energy density of metric fluctuations of Minkowaski space-time which is given by
\begin{equation}
\rho_D\sim\frac{1}{t_{p}^{2}t^2}\sim\frac{M_{p}^{2}}{t^2},
\end{equation}
where $M_p^2=(8\pi G)^{-1}$ represents the reduced Planck mass. This energy density can be viewed as the density of DE, i.e., the ADE. Following the line of the relations (1) and (2), Cai \cite{cai} proposed the original ADE model in which the energy density of the ADE has the form
\begin{equation}
\rho_D=\frac{3n^2M_{p}{^2}}{T^2},
\end{equation}
where $T$ is age of the Universe. The numerical factor $3n^2$ has been introduced to parameterize some uncertainties like the species of quantum fields in the Universe, the effects of curved space-time, etc. A number of papers on ADE models are available which deal with various aspects of the evolution of the Universe \cite{hwei,kim,neu,WEI,leme}. Wei and Cai \cite{wei} have proposed  a new version of this model referred as ``new agegraphic dark energy" model, abbreviated as NADE by replacing the cosmic age $T$ with the cosmic conformal age $\eta$ for the time scale in (3). Thus, in this model, the dark energy density is of the form
\begin{equation}
\rho_D=3n^2M_p^2\;\eta^{-2},
\end{equation}
\noindent where the conformal time $\eta$ is defined as
\begin{equation}
\eta=\int^t_0\frac{dt}{a}=\int^a_0 \frac{da}{Ha^2},
\end{equation}
\noindent where $a$ is the scale factor of the Universe and $H=\dot{a}/a$ is the Hubble parameter. Here, an overdot denotes derivative with respect to the cosmic time $t$. The most attractive merit of the NADE is that it has been proven to fit the data well. The cosmological constraints and other aspects of NADE models have been studied in the literature \cite{Wei,Setare,saaidi,Zhang}.\\
 \indent The Brans-Dicke (BD) theory \cite{brans}, which is a natural alternative and simple generalization of general relativity, provides a suitable framework for dynamical DE models. In this theory, the gravitational constant $G$ is replaced with the inverse of a time-dependent BD scalar field $\phi$, namely, $\phi=(8\pi G)^{-1}$ which couples to gravity with a coupling parameter $w$.  One important property of BD theory is that it gives simple expanding solutions for scalar field and scale factor which are compatible with the solar system observations. The BD theory explains the recent accelerated expansion of the Universe and accommodates the observational data as well \cite{bert,KIM,clif}.\\
 \indent In the BD theory, the NADE models have been studied to discuss the recent accelerated expansion of the Universe \cite{shey,ahmd,pas,fay}. To the best of our knowledge in all the papers available in the literature on HDE, ADE and NADE models in the framework of BD theory, it has been assumed that the BD scalar field evolves as a power law of scale factor, $\phi\propto a^m$, $m$ is a constant. In our previous paper \cite{singh}, we have discussed the drawback of this relation of BD scalar field in BD theory. One can observe that this relation would always give the constant deceleration parameter (DP) irrespective of the matter content. This means that the phase transition of the Universe can not be obtained with the power-law form of BD scalar field. Therefore, to over come from this problem we have proposed a new relation between BD scalar field and scale factor, namely, a logarithmic form $\phi\propto ln(\alpha+\beta a)$ which is free from this problem. In the paper we have demonstrated that the logarithmic form of BD scalar field  is not only suitable to explain the phase transition of the Universe but also it resolves the cosmic coincidence problem in an effective way. \\
 \indent As the logarithmic form of BD scalar field presents some interesting results in HDE model, therefore, it is worthwhile to discuss the evolution of the NADE model with this form. In this paper, we extend our previous work to analyze the evolution of the Universe in the NADE model in BD theory with the logarithmic form of BD scalar field. We derive the equation of state (EoS) parameter $w_D$ and DP $q$ of NADE model to discuss the evolution of the Universe. We further extend NADE model by including the interaction between dark matter and NADE in BD theory. We show that the interacting NADE model resolves the cosmic coincidence problem as the energy density ratio may evolve sufficiently slow at present.\\
 \indent The paper is organized as follows: In section 2, we present the equations of motion of NADE model in the framework of BD theory. Further, section 2 is divided into two subsections 2.1 and 2.2. In subsection 2.1, we discuss the non-interacting NADE model. Subsection 2.2 is devoted to interacting NADE model. The summary of our findings is presented in section 3.
\section{NADE in BD theory with logarithmic scalar field}
 \noindent The action for the BD theory in the Jordan frame is given by
 \begin{equation}
 S=\int d^{4}x\sqrt{-g}\left[\frac{1}{2}(-\phi R+\frac{\omega}{\phi}g^{\mu\nu}\partial_{\mu}\phi\; \partial_{\nu}\phi)+\mathcal{L}_{m}\right],
\end{equation}
 where $R$, $g$ and $\mathcal{L}_{m}$ represent the Ricci scalar curvature, determinant of the metric tensor $g_{\mu\nu}$ and the matter Lagrangian density, respectively. Here, $\phi$ and $\omega$ denote BD scalar field and BD coupling parameter, respectively. In this theory, the Newtonian constant $G$ is replaced with a time dependent scalar field $\phi$  such that $\phi=(8\pi G)^{-1}$. The dimensionless parameter $\omega$ is introduced to represent a coupling between scalar field and gravity. In Jordan frame, the scalar field and matter field do not interact and matter minimally couples to the metric.

\indent We assume a homogeneous and isotropic FRW space-time to discuss the evolution of the Universe. The FRW line element is given by  \\
  \begin{equation}
ds^{2}=dt^{2}-a^{2}(t)\left[\frac{dr^{2}}{1-k r^2}+r^{2}(d\theta^{2}+\sin^2\theta\; d\phi^2)\right],
\end{equation}
where $a$ is the cosmic scale factor and $k$ is curvature parameter. The curvature parameter has three possible values $k=-1, 0, 1$ which represent to open, flat and closed geometry of the Universe, respectively. We assume that the Universe is filled with pressureless dark matter (DM) and NADE. We exclude baryonic matter and radiation due to their negligible contribution to total energy budget in the late time evolution. \\
\indent The variation of the action (6) with respect to the metric tensor, $g_{\mu\nu}$ for the line element (7) yield the following field equations.
\begin{equation}
H^2+\frac{k}{a^2}+H\frac{\dot\phi}{\phi}-\frac{\omega}{6}\frac{\dot\phi\;^2}{\phi\;^2}=\frac{\rho_{m}+\rho_{D}}{3\phi},
\end{equation}
\begin{equation}
2\frac{\ddot a }{a}+H^2+\frac{k}{a^2}+2 H\frac{\dot\phi}{\phi}+\frac{\omega}{2}\frac{\dot\phi\;^2} {\phi\;^2}+\frac{\ddot\phi}{\phi}=-\frac{p_{D}}{\phi},
\end{equation}
\begin{equation}
\ddot\phi+3H\dot\phi=\frac{\rho_{m}+\rho_{D}-3p_{D}}{2\omega+3},
\end{equation}
where  $\rho_{m}$ and $\rho_{D}$ denote the energy density of DM and NADE respectively, and $p_{D}$ is the pressure of NADE. The Eq. (10) is known as the wave equation of the BD scalar field.\\
\indent Let us assume the relation between BD scalar field and scale factor as \cite{singh}
\begin{equation}
\phi =\phi_0\; ln(\alpha+\beta a),
\end{equation}
where $\phi_0$, $\alpha>1$ and $\beta>0$ are constants. Substituting (11) into (8) we find
\begin{equation}
H^2\left(1+\frac{k}{a^2}+\frac{\beta a}{(\alpha+\beta a)\;ln(\alpha+\beta a)}-\frac{w}{6}\frac{\beta^2 a^2}{(\alpha+\beta a)^2\;\{ln(\alpha+\beta a)\}^2}\right)=\frac{\rho_m+\rho_D}{3\phi}.
\end{equation}
It is easy to observe from the Friedmann equation (12) in BD theory that the standard cosmology of GR will be recovered in the limit of $\beta \rightarrow 0$.\\
\indent In BD theory, the BD scalar field is taken as $\phi=(8\pi G)^{-1}$ which implies $\phi=M_{p}^2$. Therefore, in this theory the energy density of NADE in (4) is now becomes
\begin{equation}
\rho_D=3n^2\phi\;\eta^{-2}.
\end{equation}
The critical energy density $\rho_{cr}$ and energy density of the curvature $\rho_k$ can be defined as follows
\begin{equation}
\rho_{cr}=3\phi H^2, \hspace{1.5cm} \rho_k=3k\phi a^{-2}.
\end{equation}
 It has been observed that the representation of quantities and equations in the terms of fractional energy densities are useful from the point of view of calculation as well as from the point of view of physical interpretation of the results. The fractional energy densities can be defined in their usual forms as
\begin{equation}
\Omega_m=\frac{\rho_m}{\rho_{cr}}=\frac{\rho_m}{3\phi H^2},
\end{equation}
\begin{equation}
\Omega_D=\frac{\rho_D}{\rho_{cr}}=\frac{n^2}{\eta^2 H^2},
\end{equation}
\begin{equation}
\Omega_k=\frac{\rho_k}{\rho_{cr}}=\frac{k}{a^2 H^2}.
\end{equation}
In what follows, we consider the non-interacting and interacting NADE models, and discuss their physical behaviors.
\subsection{Non-interacting NADE model}
 Let us assume that DM and NADE do not interact and their energy densities follow the conservation equations
\begin{equation}
\dot\rho_{m}+3H\rho_{m}=0,
\end{equation}
\begin{equation}
\dot\rho_{D}+3H(1+w_D)\rho_{D}=0,
\end{equation}
where $w_D=p_D/\rho_D$ denotes the equation of state (EoS) parameter of NADE. Using (11), (13) and (16) in the conservation equation (19), one can find the value of $w_D$ as
\begin{equation}
w_D=-1-\frac{\beta a}{3(\alpha+\beta a)\;ln(\alpha+\beta a)}+\frac{2}{3n a}\sqrt{\Omega_D}.
\end{equation}
It is to be noted that the EoS parameter $w_D$ of NADE given by (20) reduces to its respective form in GR \cite{hwei} for $\beta=0$ , that is,
\begin{equation}
w_D=-1+\frac{2}{3na}\sqrt{\Omega_D}.
\end{equation}
 From (21) we observe that the $w_D$ does not cross the phantom divide line $w_D=-1$ in the framework of GR.\\
\indent Let us examine the behavior of the term $\frac{\beta a}{(\alpha+\beta a)\;ln(\alpha+\beta a)}$ present in (20) to discuss the behavior of $w_D$. We observe that the term $\frac{\beta a}{(\alpha+\beta a)\;ln(\alpha+\beta a)}$ is zero at $a=0$, achieves its maximum value during the evolution, and converges to zero as $a\rightarrow \infty$ in the late time evolution . The maximum value of the term $\frac{\beta a}{(\alpha+\beta a)\;ln(\alpha+\beta a)}$ lies in the interval ]0, 1[ depending on the value of $\alpha$. We observe that max$\{\frac{\beta a}{(\alpha+\beta a)ln(\alpha+\beta a)}\}\rightarrow 1$ as $\alpha\rightarrow 1$ and for large value of $\alpha$ we get a small maximum value of the term. It is to be noted that the maximum value does not depend on the value of $\beta$. The time when the term will attain its maximum value depends on the values of $\alpha$ and $\beta$. For a large value of $\alpha$, it occurs in the late time and for a large value of $\beta$, it will occur in the early time, and vice-versa.\\
 \indent Now, from (20) we observe that $w_D \rightarrow -1$ in the late time as the last two terms converge to zero, i.e., the NADE mimics the cosmological constant in the late time evolution. Wei and Cai \cite{hwei} have also observed in GR that NADE mimics cosmological constant in the late time evolution. Indeed, the NADE with logarithmic form of BD scalar field in the BD theory behaves like NADE in GR in the late time evolution due to the term $\frac{\beta a}{(\alpha+\beta a)ln(\alpha+\beta a)}$ which converses to zero as $a\rightarrow \infty$. However, in the power law form of BD scalar field the NADE crosses the phantom divide line in the late time evolution because the second term is a constant $\frac{2m}{3}$ for power-law form $\phi\propto a^m$ \cite{shey}. It is also interesting to note that $w_D$ may cross the phantom divide line if ${\sqrt{\Omega_D}}<[\frac{\beta n a^2}{2(\alpha+\beta a)\;ln(\alpha+\beta a)}]$ during the evolution but ultimately it will mimic the cosmological constant in the late time evolution. However, for a sufficiently large value of $\alpha$ and a sufficiently small value of $n$ the phantom divide line crossing is not possible. If $\alpha$ has a small value ($\alpha\rightarrow 1$) and $n$ a large value then the value of $\beta$ is crucial to observe the phantom divide crossing and therefore for sufficiently small value of $\beta$ the possibility of $w_D<-1$ may be observed.\\
\indent Let us discuss the evolution of the Universe using the deceleration parameter which is defined by $q=-\frac{a\ddot a}{\dot a^2}$. Using (11), (13), (16) and (17) in (9), the deceleration parameter $q$ can be obtained as
\begin{equation}
q=\frac{1+3 w_{D}\Omega_D+\Omega_k-\frac{\beta^2a^2}{(\alpha+\beta a)^2\;ln(\alpha+\beta a)}+\frac{2\beta a}{(\alpha+\beta a)\;ln(\alpha+\beta a)}+\frac{\omega\beta^2 a^2}{2(\alpha+\beta a)^2\;\{ln(\alpha+\beta a)\}^2}}{2+\frac{\beta a}{(\alpha+\beta a)\;ln(\alpha+\beta a)}}.
\end{equation}
The value of BD parameter $\omega$ plays an important role in the value of $q$. The solar system experiment Cassini gave a very stringent high bound result on $w$ as  $|\omega|>40000$ \cite{berto,will}. However, the cosmological observations put relatively lower bounds on $\omega$ \cite{aqua,wu,cli}. The term $\frac{\beta^2 a^2}{(\alpha+\beta a)^2\;ln(\alpha+\beta a)}$ present in (22) is zero at $a=0$ which achieves its maximum value during the evolution and converges to zero in the late time evolution like the term $\frac{\beta a}{(\alpha+\beta a)\;ln(\alpha+\beta a)}$. The maximum value of this term lies in the interval ]0, 0.41[ depending on the value of $\alpha$. Due to a large value of $\omega$ suggested by the observations, the last term of the numerator in (22) containing $\omega$ will dominate in the early phase of the evolution of the Universe. Thus, we observe a positive value of $q$, i.e., the Universe evolves under the decelerated rate of the expansion. It means that the matter dominates the early phase of the expansion of the Universe. Thus, the NADE model with logarithmic form $\phi\propto ln(\alpha+\beta a)$ explains the matter dominated phase of the Universe. Since, in the late time evolution $w_D\rightarrow-1$, and last three terms of numerator and last term of denominator of (22) converge to zero, therefore, we obtain the late time value of $q$ as
\begin{equation}
q\approx\frac{1-3\Omega_D+\Omega_k}{2}.
\end{equation}
It is expected that $\Omega_D\rightarrow 1$ in the late time evolution. There are three possibilities for $\Omega_k$: (i) $\Omega_k>0$ for close Universe ($k>0$), (ii) $\Omega_k=0$ for flat Universe ($k=0$), and (iii) $\Omega_k<0$ for open Universe ($k<0$). However, the observations suggest that our Universe is almost flat ($k\approx0$) at present time. Therefore, if we take $k=0$, i.e., $\Omega_k=0$ for late time evolution then we have $q\approx\frac{1-3\Omega_D}{2}$. Thus, we may obtain an accelerated expansion for $\Omega_D>1/3$. If we take present value $\Omega_D=0.70$ obtained by resent observations then we get $q=-0.55$ which shows the accelerated expansion. One can observe that the accelerated expansion occurs more easily if we consider the open geometry of the Universe ($\Omega_k<0$). In the case of close Universe ($\Omega_k>0$), the accelerated expansion is possible but a relatively large value of $\Omega_D$ is required, i.e, a large value of $n^2$ is required to achieve it.\\
 \indent So far, we have obtained decelerated expansion in the early phase of the evolution which explains the matter dominated phase and accelerated expansion in the late time evolution which explains the NADE dominated phase. Thus, the present model successfully explains the recent phase transition of the Universe from decelerated expansion to accelerated expansion. The time of phase transition depends on the values of the parameters of the model $\alpha$, $\beta$, $n^2$ and $\omega$. It may be observed from (22) that if phase transition occurs at time $t_1$ for value $\omega_1$ then it will happen at time $t_2$ for $\omega_2$ where $t_1<t_2$ and $\omega_1<\omega_2$. For a large value of $n^2$ it occurs at earlier stage. In the similar way, one can see the effects of $\alpha$ and $\beta$ depending on their effects on the terms $\frac{\beta a}{(\alpha+\beta a)\;ln(\alpha+\beta a)}$ and $\frac{\beta^2 a^2}{(\alpha+\beta a)^2\;ln(\alpha+\beta a)}$ as discussed above.\\
  \indent The logarithmic form in NADE model is free from the constant deceleration parameter problem of the power-law form and explains the phase transition of the Universe in the framework of the BD theory. Thus, this form serves our propose well.\\
\indent Let us discuss the cosmic coincidence problem which has not been addressed for NADE models in the framework of BD theory. Steinhardt \cite{stein} addressed first time the coincidence problem- why are the matter and DE densities comparable at present time? In the conventional $\Lambda$CDM model very specific conditions of very early Universe are required to observe the matter and DE densities of the same order today. There are two ways to resolve the coincidence problem. In one of the ways, we should observe the energy density ratio $r_0\sim \mathcal{O}(1)$ for a wide range of the initial conditions. In the other way, either $r$ should converse to a constant value or evolve very slowly in the late time evolution(including present time) such that $r_0\sim \mathcal{O}(1)$. In the $\Lambda$CDM model, the energy density ratio $r$ evolves as $|\frac{\dot r}{r}|_0=3H_0$. Here and throughout the paper, the zero subscript denotes the present value of the quantity. Using (18) and (19) one can obtain the evolution of $r$ as
\begin{equation}
\dot r=3rHw_D.
\end{equation}
 As we have observed that NADE mimics cosmological constant in the late time when $a\rightarrow \infty$. Therefore, if we take $w_{D0}=-1$, we get $|\frac{\dot r}{r}|_0=3H_0$ which is same as in the $\Lambda$CDM model. Thus, there is no reduction in the acuteness of coincidence problem. However, we have a time-dependent $w_D$, therefore, it is possible to have a less acute coincidence problem if we have $w_{D0}>-1$. The quintessence like EoS ($w_D>-1$) may be achieved for $\sqrt{\Omega_D}>\frac{\beta n a^2}{2(\alpha+\beta a)\;ln(\alpha+\beta a)}$ as can be observed from (20). It is to be noted that Eq. (20) has a time-dependent second term which converses to zero in the late time whereas the second term is a constant in case of power-law form. Therefore, the quintessence like EoS may be achieved more easily with logarithmic form in comparison to the power-law form of BD scalar field in the present time. Thus, the logarithmic form is more suitable to achieve a less acute coincidence problem. The observations also suggest quintessence like EoS of NADE \cite{Wei}. Let us assume $w_{D0}=-2/3$, we obtain $|\frac{\dot r}{r}|_0=2H_0$ which is less acute coincidence problem in comparison to the $\Lambda$CDM model. On the other hand, the problem is more acute if $w_D$ has phantom like EoS ($w_D<-1$). In this case, the coincidence problem becomes more worse than the $\Lambda$CDM model. \\
  \indent In conclusion, the coincidence problem may not be resolved significantly for non-interacting NADE model in the framework of BD theory, however, there are possibilities to achieve a less acute problem. The logarithmic form of BD scalar field is more suitable to have a less acute coincidence problem than power-law form.
 \subsection{Interacting NADE model}
 A number of authors \cite{amen,zim,farr,gum,skor} have studied the interaction between DM and DE and recent cosmological observations \cite{oliv,FENG,cao,cos} also support it.  The interacting DE models  have also been studied as a possible candidate to resolve the coincidence problem \cite{mich,Campo,CAMPO}. Therefore, in this subsection we extend the NADE model by considering the interaction between DM and NADE. \\
 \indent Assuming that the DM and NADE exchange energy through the interaction $Q$, the continuity equations become
\begin{equation}
\dot\rho_{m}+3H\rho_{m}=Q,
\end{equation}
\begin{equation}
\dot\rho_{D}+3(1+w_D)H\rho_D=-Q.
\end{equation}
The sign of interaction term $Q$ is crucial and defines the direction of the energy transfer, i.e., for $Q>0$, there is an energy transfer from NADE to DM, and for $Q<0$, there is an energy transfer from DM to NADE. The interaction between DM and NADE is a quantum scale phenomenon, therefore, the interaction term $Q$ should be defined by the theory of quantum gravity. We do not have a satisfactory theory of quantum gravity, therefore, the dimension of the conservation equation has been used to choose a suitable form of $Q$ in the literature. The conservation equations of DM and NADE imply that the interaction term should has the units of energy density $\times$ $(time)^{-1}$. Therefore, we may have the interaction term of the following forms: (i) $Q\propto H \rho_D$, (ii) $Q\propto H \rho_m$ or (iii) $Q\propto H (\rho_m+\rho_D)$. In this paper, we use the last assumption which is more general than first two. This form can be rewritten as
 \begin{equation}
 Q=3b^2 H (\rho_m+\rho_D),
 \end{equation}
 where $b^2$ is a coupling constant. Using (11), (13), (16) and (27) in (26), we get
 \begin{equation}
w_D=-1-b^2(1+r)-\frac{\beta a}{3(\alpha+\beta a)\;ln(\alpha+\beta a)}+\frac{2}{3n a}\sqrt{\Omega_D}.
\end{equation}
It is easy to observe that the EoS parameter of NADE achieves its respective form in GR for $\beta=0$. We observe that $w_D$ contains the third term as time-dependent whereas it is constant term in this place in power-law form of BD scalar field. Therefore, the evolution of $w_D$ in our model will be different from that of power-law form. As discussed in section 2.1, analyzing the terms in $w_D$ in (28), one can observe that $w_D$ will definitely cross the phantom divide in the late time evolution, where $w_D\rightarrow-1-b^2(1+r)$ as $a\rightarrow\infty$. The interaction between DM and NADE is the deriving force behind the phantom crossing in the late time. The late time value of $w_D$ depends on the values of coupling constant $b^2$ and density ratio $r$. It means that it depends on how DM and NADE interact and what is the ratio of their densities. In the interacting NADE model, we observe that $w_D$ may cross the phantom divide line more easily and earlier in the framework of BD theory than GR.\\
 \indent The expression of $q$ remains same as in the case of non-interacting NADE which is given by (22). However, the evolution of $q$ is different due to different value of $w_D$ for interacting NADE given by (28). From (20) and (28), we observe that the EoS parameter of interacting NADE has an extra term $b^2(1+r)$ with negative sign than non-interacting NADE. Therefore, it is easy to observe from (22) that the recent phase transition of the Universe from matter dominated phase to NADE dominated phase is more easily achieved and we may achieve it at earlier stage in the interacting case. Thus, the interacting NADE model explains the recent phase transition of evolution of the Universe in the framework BD theory with logarithmic form of BD scalar field. \\
\indent Let us study the evolution of $r$ to check whether the interaction between DM and NADE is able to alleviate the coincidence problem or not. Using (25), (26) and (27), the evolution of $r$ can be obtained as
\begin{equation}
\dot r=3rH\left[w_{D}+b^2\frac{(1+r)^2}{r}\right].
\end{equation}
  Here, we have the sum of the terms $w_D$ and $b^2\frac{(1+r)^2}{r}$ which have negative and positive value respectively. Thus, the resultant value of terms within bracket will be less than $|w_D|$. Moreover, for a suitable value of $b^2$, we may achieve $|w_{D}+b^2\frac{(1+r)^2}{r}|<<|w_D|$. Now, it is evident from (24) and (29) that the energy density ratio $r$ may evolve more slowly in interacting NADE model than the $\Lambda$CDM model and non-interacting NADE model. Therefore, the interaction between dark components of the Universe may play an important role in the explanation of the coincidence problem. To achieve a soft coincidence we should observe $|\frac{\dot r}{r}|_0\leq H_0$ \cite{campo}. The condition of soft coincidence may be easily achieved in our model if $b^2$ satisfies the condition, $b^2\leq\frac{(1-3w_{D0})r_0}{3(1+r_0)^2}$. If we use the present observational values $r_0\simeq3/7$ and $w_{D0}\simeq-1$, we obtain the constraint $b^2\leq7/25$ to achieve soft coincidence problem. Even for a suitably  small values of $b^2$, $r$ may evolve more slowly at present which may resolve the coincidence problem more effectively. For example, if we take $b^2=\frac{1}{5}$ and $w_{D0}=-1$ then we obtain $|\frac{\dot r}{r}|_0\leq \frac{H_0}{7}$ which shows significantly slow variation of $r$ compare to conventional $\Lambda$CDM model and condition mentioned above for soft coincidence.\\
\indent Feng et al. \cite{feng} found that a small coupling constant $b^2$ is compatible with observation and also required to alleviate the coincidence problem. It also required to satisfy the second law of thermodynamics (also, see references within \cite{feng}). We also observe that the coupling constant $b^2$ should be a small quantity to avoid the coincidence problem which shows the compatibility with the observations. Thus, the interaction between DM and NADE in the framework of BD theory with logarithmic BD scalar field may resolve the coincidence problem effectively.
\section{Conclusion}
 \indent In this paper, we have investigated how the NADE model evolves in BD theory by assuming the logarithmic form of BD scalar field. The motive is that the existing power-law form of BD scalar field, $\phi=\phi_0\;a^m$ gives a constant value of DP whatever may be the energy content. Therefore, it can not describe the phase transition of the Universe. However, many authors have found time-dependent DP by using the same power-law form which shows a contradiction. To overcome from this problem we have recently proposed and analyzed the logarithmic form of BD scalar field, $\phi=\phi_0 \;ln(\alpha+\beta a)$ in HDE model (see, Ref.\cite{singh}). This form of BD scalar field is free from the above described problem and plays an important role in explaining the evolution of the Universe. It also resolves the cosmic coincidence problem. Therefore, analyzing the importance of this form, we have extended our work to NADE model within the framework of FRW line element in BD theory. We have discussed the dynamics of non-interacting and interacting NADE models by EoS parameter $w_D$ and DP $q$. We have also discussed the cosmic coincidence problem. The results of these two models are summarized as follows:\\
 \indent In non-interacting NADE model, we have observed that the NADE model successfully explains the recent phase transition from matter dominated to NADE dominated phase. In this model the NADE mimics cosmological constant in the late time whereas it achieves phantom like behavior for power-law form \cite{shey}. It has also been observed that $w_D$ may cross the phantom divide line under certain constraint but ultimately it will mimic the cosmological constant in late time evolution. Further, we have shown that a less acute coincidence problem than conventional $\Lambda$CDM model may be achieved, however, it is not able to alleviate the problem effectively. The logarithmic form of BD scalar field is more suitable to achieve a less acute coincidence problem than power-law form.\\
 \indent In interacting NADE model, the recent phase transition from decelerated phase to accelerated phase has also been achieved. However, the accelerated expansion may be achieved earlier in this model than non-interacting model. In the late time evolution, it has been analyzed that $w_D$ definitely crosses the phantom divide line. We have shown that the energy density ratio $r$ may evolve sufficiently slow in the present time which resolves the coincidence problem. Thus, the interacting NADE model may be able to alleviate the coincidence problem effectively.\\
\indent  In conclusion, we can say that the logarithmic form of BD scalar field is suitable to explain the recent accelerated expansion of the Universe in NADE model. This model shows the phase transition of the evolution of the Universe as we get the time dependent DP. It also resolves the cosmic coincidence problem effectively. Also, in the limit $\beta \rightarrow 0$, the NADE model in general relativity is recovered. \\

\end{document}